\documentclass[letterpaper,fleqn]{cas-dc}

\usepackage[numbers,sort&compress]{natbib}
\usepackage[utf8]{inputenc}
\usepackage{pdfpages}
\usepackage{verbatim}

\usepackage{datetime2}
\usepackage{datetime2-calc}
\usepackage{amsmath}
\usepackage{amssymb}

\usepackage{graphicx}
\usepackage{IEEEtrantools}

\def\ka{\kappa}

\def\ta{\tau}

\def\ch{\chi}

\def\cF{{\cal F}}

\def\cI{{\cal I}}

\def\cP{{\cal P}}

\def\cT{{\cal T}}
\def\cU{{\cal U}}

\def\dV{\dot{V}}
\def\dv{\dot{v}}
\def\du{\dot{u}}

\def\dS{\dot{S}}

\def\ddv{\ddot{v}}
\def\ddu{\ddot{u}}

\def\ddS{\ddot{S}}

\def\tcU{\tilde{\cal U}}
\def\tcF{\tilde{\cal F}}
\def\tcP{\tilde{\cal P}}
\def\fr#1#2{\dfrac{#1}{#2}}

\def\frac#1#2{{\textstyle{{#1}\over {#2}}}}
\def\te{\text}

\def\lsim{\mathrel{\rlap{\lower4pt\hbox{\hskip1pt$\sim$}}
    \raise1pt\hbox{$<$}}}
\def\gsim{\mathrel{\rlap{\lower4pt\hbox{\hskip1pt$\sim$}}
    \raise1pt\hbox{$>$}}}

\def\lrpartial{\raise 1pt\hbox{$\stackrel\leftrightarrow\partial$}}

\newcommand{\beq}{\begin{equation}}
\newcommand{\eeq}{\end{equation}}
\newcommand{\bea}{\begin{eqnarray}}
\newcommand{\eea}{\end{eqnarray}}
\newcommand{\rf}[1]{(\ref{#1})}
\def\tsc#1{\csdef{#1}{\textsc{\lowercase{#1}}\xspace}}
\tsc{WGM}
\tsc{QE}
\tsc{EP} 
\tsc{PMS}
\tsc{BEC}
\tsc{DE}

\begin{document}
\let\WriteBookmarks\relax
\def\floatpagepagefraction{1}
\def\textpagefraction{.001}
\shorttitle{Observables on a collapsing black hole background}
\shortauthors{M.S. Berger and Z. Liu}

\title [mode = title]{Observables in quantum field theory on a collapsing black hole background}                      
\author{Micheal S. Berger}[orcid = 0000-0003-1705-3179]

\ead{berger@indiana.edu}

\author{Zhi Liu}[orcid = 0000-0002-8716-8097]
\cormark[1]
\ead{liu333@iu.edu}

\address{Physics Department, Indiana University,
Bloomington, IN 47405, USA}

\cortext[cor1]{Corresponding author}

\begin{abstract}
We study the quantum radiation from a curved background using scalar observables constructed from the (1+1) dimensional renormalized stress-energy tensor (RSET). We compute energy density $\cU$, flux $\cF$ and pressure $\cP$ for an arbitrary collapse scenario and an arbitrary observer. The results show a clear structure: $\cF - \cU$ represents the collapse-independent ingoing modes, $\cF + \cU$ represents the collapse-dependent outgoing modes and $\cP - \cU$ represents the trace anomaly. We also compute the observables constructed from the perception renormalized stress-energy tensor (PeRSET). We find they have the same collapse-dependent contribution comparing to their RSET-related counterparts. For free-falling observers, in particular, the PeRSET-related observables are the same as the corresponding RSET-related ones, except the energy densities differ by the trace anomaly.
\end{abstract}

\begin{highlights}
\item Observables separate into ingoing modes, outgoing modes and the trace anomaly
\item Collapse-dependent contributions are the Schwarzian derivative of the scaling factor
\item Perceived flux and pressure same as conventional ones for free-falling observers
\end{highlights}


\maketitle

\section{Introduction}
\label{se:intro}

Quantum field theory in curved spacetime has made famous theoretical predictions such as Hawking radiation \cite{Hawking1974, Hawking1975} and Unruh effect \cite{Fulling1973, Davies1975, Unruh1976}. One approach to investigate these effects is by physical observables constructed from the stress-energy tensor, such as energy density and flux (see e.g. \cite{Ford1993,Smerlak2013,Eune2014,Singh2014,Chakraborty2015,Firouzjaee2015,Kim2015,Gim2016,Dey2017,Dey2019} taking this approach); another approach is to measure the number of particles detected by a particle detector. With the exact form of the renormalized stress-energy tensor (RSET) in (1+1) dimensional conformal fields, several authors have evaluated the RSET or related observables for concrete collapse scenarios \cite{Smerlak2013, Singh2014, Chakraborty2015,Paranjape2009, Banerjee2009, Dai2016a, Dai2016b, Barcelo2019}. One question is can we generalize the calculation to an arbitrary collapse scenario and for an observer taking an arbitrary trajectory? We can then analyze the structure of the result and identify or interpret the different contributions, which can give us a deeper understanding of the radiation in curved background.

The RSET has tensorial nature and the related observables are scalars under coordinate transformations. However, these observables only depend on the observer's velocities but not on the accelerations. So, the first approach does not account for the Unruh effect. This motivates the introduction of the perception renormalized stress-energy tensor (PeRSET) \cite{Barbado2016}, which has the tensorial nature and takes the observer's acceleration into account. Observables constructed from this tensor are therefore closer to what a particle detector would see. One natural thing to do is to make the similar generalization to arbitrary collapse scenarios and arbitrary observers to better understand PeRSET and its connection to RSET.

The structure of this paper is as follows: in section 2, we set up the preliminaries and describe a general collapse geometry using the scaling factor; in section 3, we express the RSET and the scalar observables in terms of the scaling factor and discuss the properties of these observables; in section 4 we investigate the PeRSET-related observables and compare with the RSET-related counterparts; we summarize in section 5.

\section{Quantum fields and background geometry}
\label{se:coord}
We consider the massless scalar field with minimal coupling, make the $s$-wave approximation and ignore the effective potential term in the wave equation, which is effectively a (1+1) dimensional field theory. This allows us to compute the RSET exactly.

\begin{figure}[t!]
\centerline{\includegraphics[width=0.7\columnwidth]{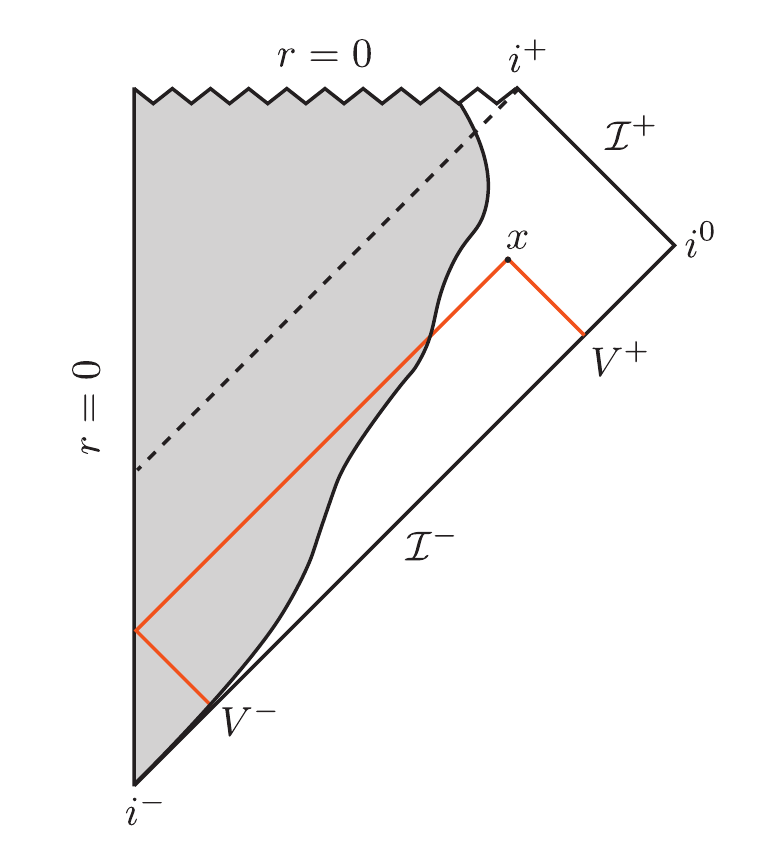}}
\caption{The Penrose diagram of an arbitrary collapsing scenario forming a black hole. The shaded region is inside the classical matter. For a given point $x$, the global null coordinates ($V^-, V^+$) are constructed by following the outgoing or ingoing null rays back to $\cI^-$, respectively.}
\label{Penrose}
\end{figure}

We assume the background geometry to be spherically symmetric but otherwise arbitrary and use the units in which the Schwarzschild radius $r_s= 2M = 1$. When written in Eddington-Finkelstein coordinates $(u, v)$ ($u = t - r^*$, $ v = t + r^*$ and $r^*= r + \ln|r - 1|$ both inside and outside the horizon), the metric outside the classical matter is
\beq \label{1}
ds^2_{\te{ext}} = -\left(1- \fr {1} {r}\right)dudv.
\eeq
Another set of null coordinates ($V^-, V^+$) \cite{Smerlak2013, Singh2014, Chakraborty2015, Barcelo2008}, known as the global null coordinates, are used to define the ``in'' vacuum, in which observers at the past null infinity $\cI^-$ see no particles. In these coordinates, $V^+$ is the same as $v$ and $V^-$ is the $v$ coordinate when tracing the outgoing null ray back to $\cI^-$ (Fig.~\ref{Penrose}). To rewrite the metric in these coordinates, we define the scaling factor $S$ as
\beq
S \equiv \dfrac {du} {dV^-},
\eeq
which is the same as $1/\dot{p}(u)$ in Refs.~\cite{Barcelo2008, Dey2017}. 

Physically, $S$ represents the amount of redshift when a light ray coming from $\cI^-$, passes through the collapsing matter and finally reaches the future null infinity $\cI^+$. If the classical matter is static, the amount of blueshift when a light ray comes in cancels that of redshift when it comes out and $S = 1$. If the matter is collapsing monotonically, the redshift exceeds the blueshift, because the matter is more compact when the light rays come out \cite{Barcelo2008}. Therefore $du > dV^-$ and $S > 1$.

\begin{figure}[t!]
\centering
\includegraphics[width=\columnwidth]{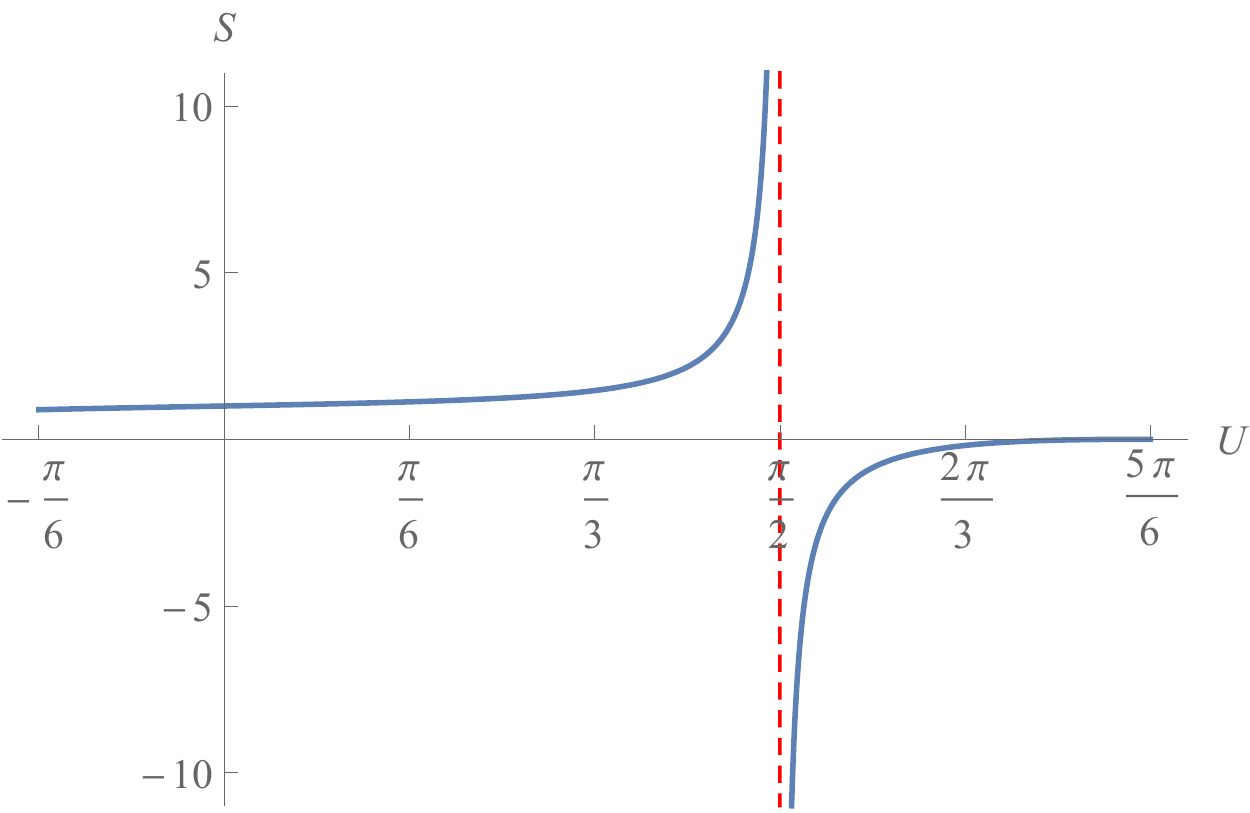}
\caption{The scaling factor $S$ for the FRW dust ball collapse for $\ch_0 = \pi/6$, which specifies the dust ball surface. $U$ is the outgoing null coordinate inside the dust ball. $S$ blows up and changes sign at the horizon ($U = \pi / 2$) due to the property of $u$ coordinate there.}
\label{PlotS}
\end{figure}

The Friedmann-Robertson-Walker (FRW) dust ball collapse considered in Ref.~\cite{Chakraborty2015} gives us more insight into the properties of $S$. In this model, the dust ball is described by the closed FRW metric with $U$ being the outgoing null coordinate. In Fig.~\ref{PlotS}, we plot $S$ as a function of $U$ for $\ch_0 = \pi / 6$, which specifies the surface of the dust ball. $U = \ch_0$ is the boundary of whether a null ray has experienced the dust ball's expanding period. This explains $S < 1$ when $U < 0$, since the null ray experiences a long enough expanding period and the blueshift dominates. We can also see $S$ goes to infinity and switches sign at $U = \pi/2$, which corresponds to the event horizon ($U = \pi - 3\ch_0$). This is because at the horizon, $u$ blows up and $du$ changes sign while $dV^-$ at this point is positive and finite. Therefore, $S$ depends on both the collapse and the nature of $u$ coordinate.

\section{RSET-related observables}

In this section, we compute the observables constructed from RSET for a general observer in an arbitrary collapsing scenario. We then discuss the properties of the results.

The RSET for a massless scalar field in a conformal metric 
\beq
ds^2 = -C(V^-,V^+)dV^-dV^+
\label{3}
\eeq
has the components \cite{Davies1976a,Davies1977a,Davies1976b,Bunch1978}
\begin{align}
&\langle T_{--}\rangle ={\dfrac 1 {12\pi}}\left [{\dfrac 1 2}{\dfrac{\partial ^2_- C} C} -{\dfrac 3 4}\left (\dfrac {\partial _-C} C\right )^2\right ] \;,
\nonumber \\
&\langle T_{++}\rangle ={\dfrac 1 {12\pi}}\left [{\dfrac1 2}{\dfrac{\partial ^2_+ C} C} -{\dfrac3 4}\left ({\dfrac{\partial _+C} C}\right )^2\right ]
\;,
\nonumber \\
&\langle T_{+-}\rangle =-{\dfrac 1 {24\pi}}\left [{\dfrac{\partial _+\partial _-C} C}-{\dfrac {\partial _+C} C}{\dfrac{\partial _-C} C} \right ]\;,
\label{emt}
\end{align}
where the vacuum is the ``in'' vacuum. These formulas are valid for the metric described in general null coordinates, not just the global null coordinates $V^\pm$ defined in the previous section.

Applying these equations to the exterior metric discussed in Section 2
\beq
ds^2_{\te{ext}} = -\left(1- \fr {1} {r}\right) SdV^-dV^+,
\label{5}
\eeq
we obtain, at radius $r$,
\begin{align}
&\langle T_{--}\rangle={\dfrac{\kappa ^2} {48\pi}}\left [\left (\dfrac 3 {r^4}-\dfrac 4 {r^3}\right )S^2
-16S^{1/2}\partial _-^2S^{-1/2}\right ]\;,
\nonumber \\
&\langle T_{++}\rangle={\dfrac{\kappa ^2} {48\pi}}\left (\dfrac 3 {r^4}-\dfrac 4 {r^3}\right )\;,
\nonumber \\
&\langle T_{+-}\rangle={\dfrac{\kappa ^2} {12\pi}}\left (\dfrac 1 {r^4}-\dfrac 1 {r^3}\right )S\;,
\label{6}
\end{align}
where $\ka = 1/2$ is the surface gravity of the horizon.

The physical observables can be constructed by contracting the RSET with the observer's velocities $u^\mu$ and the normal vectors $n^\mu$, which satisfy $u_\mu n^\mu = 0$. In the global null coordinates, these vectors are
\beq
u^\mu = (\dV^-, \dV^+), \qquad n^\mu = (-\dV^-, \dV^+),
\eeq
where a dot denotes the derivative with respect to proper time $\ta$ and we choose $n^\mu$ to be in the direction of decreasing $V^-$. The scalar observables, energy density $\cU$, flux $\cF$ and pressure $\cP$, are
\begin{align}
{\cal U}&\equiv \langle T_{\mu\nu}\rangle u^\mu u^\nu  \nonumber \\
&=\langle T_{++} \rangle (\dV^+)^2 + 2 \langle T_{+-} \rangle \dV^+ \dV^- +\langle T_{--}\rangle(\dV^-)^2\;, 
\nonumber \\
{\cal F} &\equiv -\langle T_{\mu\nu}\rangle u^\mu n^\nu \nonumber\\
&= - \langle T_{++} \rangle (\dV^+)^2  +\langle T_{--}\rangle(\dV^-)^2 \;,
\nonumber \\
{\cal P} &\equiv \langle T_{\mu\nu}\rangle n^\mu n^\nu \nonumber\\
&=\langle T_{++} \rangle (\dV^+)^2 - 2 \langle T_{+-} \rangle \dV^+ \dV^- +\langle T_{--}\rangle(\dV^-)^2 \;.
\label{8}
\end{align}

By Eqs. \rf{6} to \rf{8}, $\dV^+ = \dv$ and $\dV^- = \du S^{-1}$, we obtain
\begin{align}
\cU =& \dfrac {\ka^2}{48\pi} \Big [\left(\dfrac {3}{r^4} -\dfrac {4} {r^3}\right)(\du^2 +\dv^2) - \dfrac {8}{r^3} \nonumber\\
&- 16 S^{-3/2}\partial_-^2 S^{-1/2}\du^2\Big] \;,\nonumber\\
\cF =&  \dfrac {\ka^2}{48\pi} \left[\left(\dfrac {3}{r^4} -\dfrac {4} {r^3}\right)(\du^2 -\dv^2) - 16 S^{-3/2}\partial_-^2 S^{-1/2}\du^2\right] \;,\nonumber\\
\cP =& \dfrac {\ka^2}{48\pi} \Big[\left(\dfrac {3}{r^4} -\dfrac {4} {r^3}\right)(\du^2 +\dv^2) + \dfrac {8}{r^3} 
\nonumber \\
&- 16 S^{-3/2}\partial_-^2 S^{-1/2}\du^2\Big] \;.
\label{9}
\end{align}

To understand the structure of this result, we can consider some combinations of these observables. For example,
\beq \label{10}
\cP - \cU = -4 \langle T_{+-} \rangle \dV^+\dV^- = \dfrac{\ka^2}{3\pi r^3}.
\eeq
By Eq. \rf{3} and $d\ta^2 = -ds^2 $, we have $\dV^+ \dV^- = 1/C$. Therefore,
\beq \label{11}
\cP - \cU = -4 \langle T_{+-} \rangle C^{-1} = 2\langle T_{+-} \rangle g^{+-}= \langle T_{\mu}^{\mu}\rangle
\eeq
is the trace anomaly \cite{Capper1974, Capper1975, Deser1976, Duff1977}. We can also see this from the fraction in Eq. \rf{10}:
\beq
\cP - \cU = \dfrac{1}{24\pi} \dfrac{2}{r^3} = \dfrac{R}{24\pi} =  \langle T_{\mu}^{\mu}\rangle\;.
\eeq
Here $R = 2/r^3$ is the Ricci scalar of the two-dimensional Schwarzschild geometry and the trace anomaly of the two-dimensional scalar field is $R/(24\pi)$ \cite{Christensen1978}. From Eq. \rf{11}, we can see that this result is true for all vacuum states. So, $\cP - \cU$ shares the properties of the trace anomaly, which only depends on the local geometry and is independent of the vacuum state. 

Another combination
\begin{align}
\cF - \cU &= -2 \langle T_{++} \rangle (\dV^+)^2 -2 \langle T_{+-} \rangle \dV^+ \dV^- \nonumber \\
&=\dfrac{\ka^2}{24\pi}\left[-\left(\dfrac{3}{r^4} - \dfrac{4}{r^3} \right)\dv^2 + \dfrac{4}{r^3}\right].
\label{13}
\end{align}
By the definitions of $V^+$ and $V^-$, $\cF - \cU$ is related to the ingoing modes, which come from $\cI^-$ and arrive at the considered point $x$ without going through the classical matter. It is unrelated to the collapse detail but is associated with the observer trajectory because of the $\dv^2$ term. Similarly,
\begin{align}
\cF + \cU &=  2 \langle T_{+-} \rangle \dV^+ \dV^- + 2 \langle T_{--} \rangle (\dV^-)^2\nonumber \\
&=\dfrac{\ka^2}{24\pi}\left[\left(\dfrac{3}{r^4}-\dfrac{4}{r^3} -16 S^{-3/2}\partial_-^2 S^{-1/2}\right)\du^2 - \dfrac{4}{r^3} \right]
\label{14}
\end{align}
represents the outgoing modes starting form $\cI^-$ and passing through the collapsing matter before reaching $x$. It depends on how the collapse happened. Here, we choose to define $\cF - \cU$ and $\cF + \cU$ to be the ingoing and outgoing modes, respectively, to avoid introducing the trace anomaly as a third component, which contributes to $\cU$. One can alternatively define $-2 \langle T_{++} \rangle (\dV^+)^2$ to be the ingoing modes and $2 \langle T_{--} \rangle (\dV^-)^2$ to be the outgoing modes. From Eq. \rf{14}, we can see the dependence on the collapse is reflected in the scaling factor $S$ and is in the form of a Schwarzian derivative \cite{Barcelo2008} ($S^{1/2}\partial^2_- S^{-1/2}$ term in $\langle T_{--}\rangle$ in Eqs. \rf{6}), since the $S^2$ factor in $\langle T_{--} \rangle$ cancels with $S^{-2}$ in $(\dV^-)^2$). The Schwarzian derivative of a function $f(z)$ with respect to $z$ is defined as
\beq
\{f;z\}\equiv \dfrac{f'''}{f'}-\dfrac{3}{2}\left(\dfrac{f''}{f'} \right)^2,
\eeq
where prime denotes the derivative with respect to $z$. So,
\beq
S^{1/2}\partial^2_- S^{-1/2}= -\dfrac 1 2 \{u;V^- \}.
\eeq
The structure in these observables (Eq. \rf{9}) now becomes more apparent: the collapse-dependent terms are proportional to the Schwarzian factor $S^{-3/2}\partial^2_- S^{-1/2}$, and the remaining terms are the corresponding observables in Boulware vacuum \cite{Boulware1975}, which we call Boulware terms. The Boulware vacuum is associated with ($u, v$) coordinates. 

To better understand the physics at the horizon, we analyze the Schwarzian factor by giving an approximate expression for $S$ near the horizon. Since $S$ is a property of the collapse, $du/dV^-$ can be evaluated by approaching the considered point in any manner. Along the line of constant $v$, $u = v -2r^*$. Then, near the horizon
\beq
S = -\dfrac{2r}{r-1}\dfrac{dr}{dV^-} \sim -\dfrac{2r}{r-1} \dfrac{1-r}{V_0 - V^-} \overset{r\rightarrow 1}{\longrightarrow} \dfrac{2}{V_0 - V^-},
\eeq
where $V_0$ is the $V^-$ coordinate of the horizon and $\sim$ is the approximation of using the slope of a secant as the derivative very close to the horizon. Plugging this approximate expression into the Schwarzian factor, we find $S^{-3/2}\partial^2_- S^{-1/2}=-1/16$ at the horizon, which is first shown in Ref.~\cite{Barcelo2008}. This is related to the necessary condition for $\cF + \cU$ being finite at the horizon. We can see this by letting $r\rightarrow 1$ in Eq. \rf{14}. Since $u\rightarrow \infty$ and $d\ta$ is finite, $\du$ blows up. In order to have a finite $\cF + \cU$, the terms in the parentheses must vanish, which implies $S^{-3/2}\partial^2_- S^{-1/2} =-1/16$. Classically, the contribution from the outgoing modes at the horizon is equal to zero. A nonzero value of $\cF + \cU$ is due to quantum effects.

\begin{figure}[t!]
\centerline{\includegraphics[width=\columnwidth]{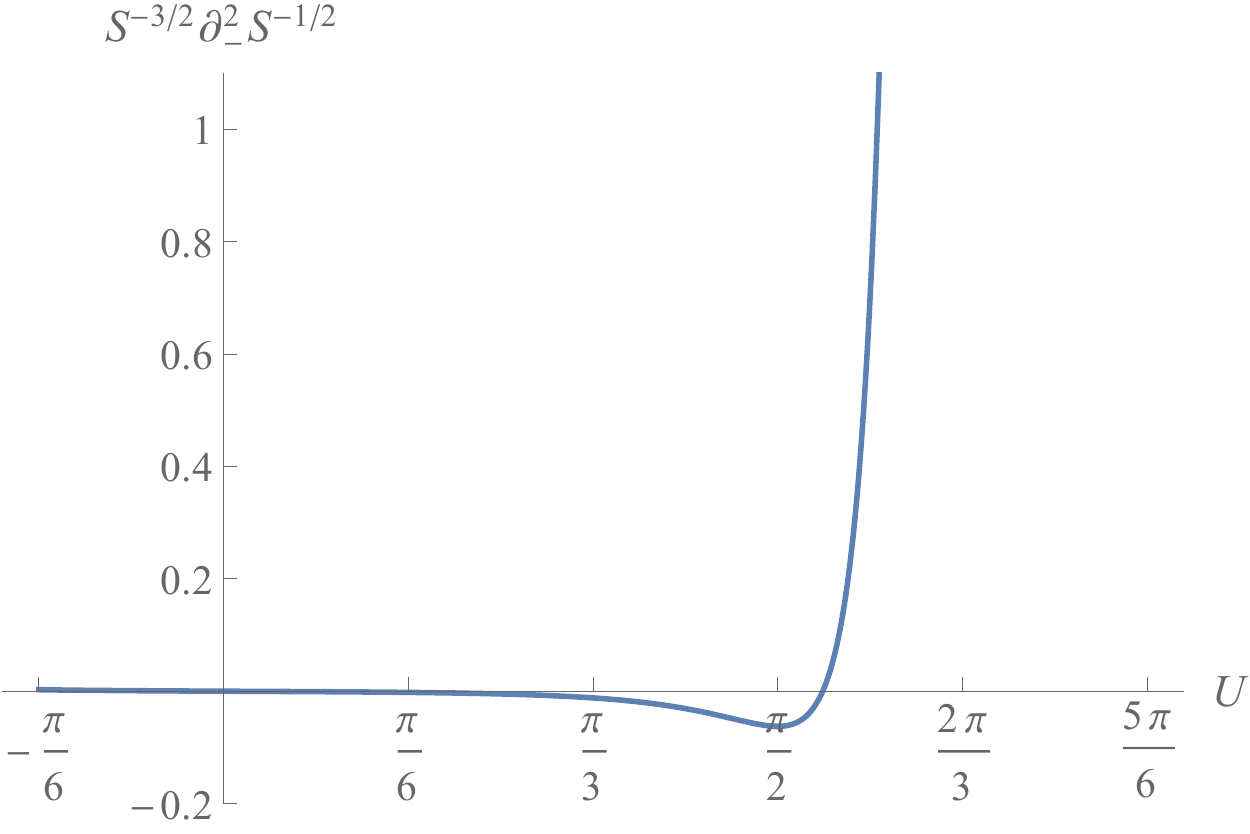}}
\caption{The Schwarzian factor for the FRW dust ball collapse for $\ch_0 = \pi/6$.
}
\label{PlotSchwarzian}
\end{figure}

We can demonstrate this property of the Schwarzian factor in the FRW dust ball model (see Section 2). Fig.~\ref{PlotSchwarzian} plots this factor as a function of $U$ for $\ch_0 = \pi /6$. The minimum is achieved at the horizon ($U = \pi/2$) with a value $-1/16$, consistent with the discussion above.

\section{PeRSET-related observables}

RSET-related observables depend on the observer's velocity, but not on the acceleration (Eqs. \rf{9}). To account for the acceleration and retain the tensorial nature, Barbado et al. \cite{Barbado2016} introduce the perception renormalized stress-energy tensor (PeRSET) as the difference between the RSET in the original vacuum and that in the observer's local vacuum $| \tilde{0}\rangle$:
\beq \label{18}
\cT_{\mu\nu} \equiv  \langle T_{\mu\nu} \rangle -\langle \tilde{0} | T_{\mu\nu} | \tilde{0}\rangle\;.
\eeq
From this definition, observers have zero perception in their local vacuums. The authors of \cite{Barbado2016} construct the coordinates associated with this vacuum using the observer's proper time and define the PeRSET-related observables $\tcU$, $\tcF$ and $\tcP$ in a similar manner. Since $\cP -\cU$, the trace anomaly, is the same for different states, we have $\tcP = \tcU$.

Applying the expressions for $\tcU$ and $\tcF$ (Eqs. (36) and (37) in Ref. \cite{Barbado2016}) to the general collapse geometry we considered, we find, for ``in'' vacuum,
\begin{align}
\tcU = \tcP =& \dfrac{1}{48\pi} \Big (-4\du^2 S^{-3/2}\partial_-^2 S^{-1/2} \nonumber \\
&+ \dfrac{\ddu^2}{\du^2} +\dfrac{\ddv^2}{\dv^2} - 2 \dfrac{d^2}{d\ta^2} \ln \du \dv \Big)\;,
\nonumber\\
\tcF =& \dfrac{1}{48\pi} \Big (-4\du^2 S^{-3/2}\partial_-^2 S^{-1/2} \nonumber \\
&+ \dfrac{\ddu^2}{\du^2} -\dfrac{\ddv^2}{\dv^2} - 2 \dfrac{d^2}{d\ta^2} \ln \dfrac{\du} {\dv} \Big)\;.
\end{align}
These equations can be derived by trading $\dS$ and $\ddS$ with $u$ derivatives and using $ S^{-1/2}\partial_u^2 S^{1/2} = - S^{-3/2}\partial_-^2 S^{-1/2} $, which comes from the inversion formula for Schwarzian derivative: $\{w;z\} = -(dw/dz)^2\{z;w\}$. We find the $S$-dependent contribution is the same as that in $\cU$, $\cF$ and $\cP$. This is because $|\tilde{0}\rangle$ is the observer's local vacuum and $\langle \tilde{0} | T_{\mu\nu} | \tilde{0}\rangle$ only depends on local property. As with RSET-related observables, the remaining terms in $\tcU$, $\tcP$ and $\tcF$ are still Boulware terms. In addition, $\tcF -\tcU$ represents the $S$-independent ingoing modes, $\tcF+\tcU$ represents the $S$-dependent outgoing modes. The identification of ingoing and outgoing modes is a property of stress-energy tensors and, therefore, applies to the PeRSET. In short, RSET- and PeRSET-related variables share the same general properties, except that the trace anomaly is absent in perceived quantities.

Some new properties appear when we consider observers taking special trajectories. For a free-falling observer starting from rest at radius $r_i$ and reaching radius $r$,
\begin{align}
\label{20}
&\cU - \tcU = -\dfrac{\ka^2}{3\pi r^3}\;, \nonumber \\
&\cP - \tcP = 0\; , \nonumber \\
&\cF - \tcF = 0 \;.
\end{align}
By Eq. \rf{18}, the difference between the RSET-related observable and its PeRSET counterpart is the observable associated with $\langle \tilde{0} | T_{\mu\nu} | \tilde{0}\rangle$, therefore relations like Eqs. \rf{20} are valid for all vacuum states. From Eqs. \rf{20}, we also see that $\langle \tilde{0} | T_{\mu\nu} | \tilde{0}\rangle$ has the properties of dust for free-falling observers, with zero pressure and flux but nonzero energy density. The energy density for $\langle \tilde{0} | T_{\mu\nu} | \tilde{0}\rangle$ is negative the trace anomaly. This is a result of a more general relation
\beq \label{21}
(\cP - \tcP) - (\cU - \tcU) = \dfrac{\ka^2}{3\pi r^3}\;, 
\eeq
which can be derived from $\cP - \cU = \ka^2/(3\pi r^3)$ and $\tcP = \tcU$. Why is this energy density nonzero? Because it is related to the RSET, and the trace anomaly doesn't allow both pressure and energy density to be zero. Since the trace anomaly is proportional to the Ricci scalar in (1+1) dimensions, $\langle \tilde{0} | T_{+-} | \tilde{0}\rangle$ is nonzero, which can contribute to the energy density. We also see from Eqs. \rf{13} and \rf{14} that the trace anomaly is contributing to $\cU$. The PeRSET, on the other hand, is traceless \cite{Barbado2016}. So it's reasonable to have different $\cU$ and $\tcU$.

For a static observer at fixed radius $r$, we also have $\tcF = \cF$. But $\cU -\tcU$ and $\cP - \tcP$, which satisfy Eq. \rf{21}, are distributed differently:
\begin{align}
\label{22}
&\cU - \tcU = \dfrac{\ka^2}{24\pi}\dfrac{r}{r-1}\left(\dfrac{7}{r^4}-\dfrac{8}{r^3}\right)\;,\nonumber\\
&\cP - \tcP = -\dfrac{\ka^2}{24\pi}\dfrac{r}{r-1}\dfrac{1}{r^4}\;.
\end{align}
This is because $\du = \dv = (r/(r-1))^{1/2}$ and $\ddu = \ddv = 0$, which give vanishing Boulware terms for $\cF$, $\tcU$, $\tcP$ and $\tcF$. Physically, in the Boulware vacuum, there is no Hawking effect for a static observer at finite $r$, since the Hawking radiation is only outgoing and is present in the ``in'' vacuum in the form of the Schwarzian factor. So $\cF$ is zero. Since the observer is accelerating relative to the local inertial frame, the traditional interpretation (using the RSET) is that Unruh effect still exists and $\cU$ and $\cP$ are nonzero. For perceived observables, there is also no Unruh effect according to the interpretation given in Refs. \cite{Barbado2016,Barbado2016b} that the Unruh effect is associated with the acceleration relative to the asymptotic region. These observables separate out the Hawking radiation and Unruh radiation as radiation action and radiation backreaction on a detector; a static observer only perceives Hawking radiation and no Unruh radiation. As we have mentioned earlier, the Hawking radiation is zero, therefore $\tcU= \tcP = \tcF =0$ in the Boulware vacuum \cite{Barbado2016}. This explains why $\cF - \tcF =0$ but $\cU - \tcU$ and $\cP -\tcP$ are nonzero. 

For completeness, the expressions of $\cU$, $\cF$ and $\cP$ for free-falling observers and static observers in the ``in'' vacuum are given in the Appendix A.

\section{Summary}
In this work, we generalize the computation of scalar physical observables, energy density $\cU$, flux $\cF$ and pressure $\cP$, to arbitrary collapse scenarios and arbitrary observers. Even though the generalization is straightforward, this abstraction allows us to unravel the structure of these observables. By using the ``in'' vacuum, we identify the collapse-independent ingoing modes $\cF - \cU$, collapse-dependent outgoing modes $\cF + \cU$ and the trace anomaly $\cP - \cU$. The collapse-dependent contributions are all in the form of a Schwarzian derivative of the scaling factor. This Schwarzian factor is equal to $-1/16$ at the horizon, which is a necessary condition for the outgoing modes $\cF + \cU$ to be finite there. The collapse-independent parts are Boulware terms (the same observable in the Boulware state).

We also compute the PeRSET-related observables in the same general situation. Compared to the RSET-related ones, they consist of the same collapse-dependent contributions and their own Boulware terms. Similar interpretations of ingoing and outgoing modes also apply. But the trace anomaly disappears since PeRSET is defined as the difference of RSET in two vacuum states. When considering special observers, we find $\tcF = \cF$ for both free-falling and static observers with $\tcP = \cP$ also applying to free-falling observers. $\cP - \tcP$ and $\cU - \tcU$ distribute in a way so that their difference equals the trace anomaly.

\appendix
\section{Observables for static and free-falling observers}
We give the expressions of $\cU$, $\cF$ and $\cP$ for static observers and free-falling observers in the ``in'' vacuum. These equations can be obtained from Eqs. \rf{9} or by relevant substitutions in expressions for the FRW dust ball collapse \cite{Chakraborty2015}.

For a static observer at radius $r$ outside the horizon, we have
\begin{align}
&\cU = \dfrac{\ka^2}{24\pi}\dfrac{r}{r-1}\Big(\dfrac {7}{r^4} -\dfrac{8}{r^3} -8S^{-3/2}\partial_-^2 S^{-1/2}\Big)\;, \nonumber \\
&\cF =  \dfrac{\ka^2}{24\pi}\dfrac{r}{r-1} (-8S^{-3/2}\partial_-^2 S^{-1/2})\;, \nonumber \\
&\cP = \dfrac{\ka^2}{24\pi}\dfrac{r}{r-1} \Big(-\dfrac{1}{r^4}-8S^{-3/2}\partial_-^2 S^{-1/2}\Big)\;.
\end{align}

For a free-falling observer starting from rest at radius $r_i$, the energy per unit mass is $E = (1- \frac{1}{r_i})^{1/2}$. 
When the observer reaches radius $r$, we have
\begin{align}
\cU=& \dfrac{\ka^2}{24\pi}\dfrac{r^2}{(r-1)^2}\Big[E^2\left(\dfrac{6}{r^4} - \dfrac{8}{r^3}\right) + \dfrac{r-1}{r^5} 
\nonumber\\
&-8\left(E + \sqrt{E^2 - \frac{r-1}{r}}\right)^2 S^{-3/2}\partial^2_- S^{-1/2}
\Big] \;,
\nonumber \\
\cF = &\dfrac{\ka^2}{24\pi}\dfrac{r^2}{(r-1)^2}\Big[E\sqrt{E^2-\frac{r-1}r} \left(\dfrac{6}{r^4} - \dfrac{8}{r^3}\right) 
\nonumber\\
&-8\left(E + \sqrt{E^2 - \frac{r-1}{r}}\right)^2 S^{-3/2}\partial^2_- S^{-1/2}\Big] \;,
\nonumber \\
\cP = & \dfrac{\ka^2}{24\pi}\dfrac{r^2}{(r-1)^2}\Big[E^2\left(\dfrac{6}{r^4} - \dfrac{8}{r^3}\right) + \dfrac{r-1}{r}\left(\dfrac{8}{r^3}-\dfrac{7}{r^4}\right) 
\nonumber\\
&-8\left(E + \sqrt{E^2 - \frac{r-1}{r}}\right)^2 S^{-3/2}\partial^2_- S^{-1/2}\Big] \;.
\end{align}

\bibliographystyle{elsarticle-num}
\bibliography{mybibtex}

\end{document}